\documentclass{article}

\begin{document}

\title{Binding in charged spherically symmetric objects}

\author{Matthew Corne \and
        Arkady Kheyfets \and
        Jennifer Piasio \\
        \small{Department of Mathematics, North Carolina State University, Raleigh, NC 27695-8205} \and
        Chad Voegele \\
        \small{Department of Mathematics, University of Virginia, Charlottesville, VA 22904-4137}}
        
\maketitle        

\begin{abstract}
We consider the subject of self--binding in static, spherically symmetric objects consisting of a 
charged fluid. We have shown previously that in the case of a perfect fluid, only the localized part 
of the mass contributes to gravitational self--binding 
of such objects and that in the limiting case of objects comprised purely of electromagnetic mass, 
there is no gravitational binding. Here, we extend this result to the more general case of an anisotropic fluid. 
Our inspection of the Oppenheimer--Volkov equation allows tracking of both gravitational and non-gravitational 
contributions to binding of spherically symmetric objects and shows that those with pure electromagnetic 
mass cannot exist.

\end{abstract}

\section{Introduction}
\label{intro}
In the following, we consider the issue of self--binding for static spherically symmetric 
systems in which the source of the gravitational field is a static spherically symmetric object formed by 
a charged perfect fluid (or its anisotropic generalization). This analysis can be applied either to charged 
spherical stars 
\cite{Ndu,KB,Sin,PS,RD} or to the Abraham--Lorentz model of an electron \cite{TRK,Gau,Bon}.  
In the latter case, the original hope was that, in general relativity, the electric field 
would contribute sufficiently 
to the gravitational binding energy to stabilize such a model without resorting to Poincar\'{e} 
stress.  However, this does not happen 
for any known solution of such a model. Our analysis of this issue reveals that it is impossible in principle for reasons which 
are entirely model--independent.  Accordingly, we have not constructed any models, nor are we 
attempting to do so.  Rather, we are interested \emph{only} in the general case of a charged, static spherically symmetric object 
and the contributions to its gravitational and total binding. 

Section 2 introduces basic notations and equations (Einstein -- Maxwell) describing static spherically symmetric 
objects consisting of a charged perfect fluid (isotropic case with radial pressure equal to tangential pressure). This description, 
although very similar and in places coinciding with that of our previous paper \cite{CKM}, is unavoidable as it defines the object of our study. 
Following this, the expressions for the total mass (Reissner--Nordstr{\o}m parameter) and the gravitational binding energy are 
presented. These expressions are based on what is called the \emph{mass function} (not to be confused with the total mass). 
The electromagnetic contribution to the total mass is written 
as an integral over the volume of the object, although it includes the electromagnetic mass outside of the body as well. 
This contribution is shown to be non--localizable and not 
to affect the gravitational binding energy. Thus, for an object to be bound 
gravitationally (and this was the principal result of \cite{CKM}), it must contain enough perfect fluid to generate binding 
sufficient to overcome electrostatic repulsion. Some 
additional aspects of issues concerning the total mass and gravitational binding are discussed (and from there on the content of the present paper does not 
intersect with \cite{CKM}), such as an alternative definition of the total mass (Tolman--Whittaker) in its 
modern version provided by Komar and what is sometimes referred to as the general relativistic version of the virial theorem, 
which states that the Tolman--Whittaker definition is equivalent to the one that is based on the mass function. The generalization 
of the expressions for the total mass and gravitational binding energy of a charged anisotropic fluid (with radial pressure unequal to 
tangential pressure) is shown to be trivial and not to change our previous conclusions.  
 
Our previous analysis of charged, static spherically symmetric objects based only on gravitational binding is incomplete. If gravitational 
binding is insufficient 
to overcome electrostatic repulsion, it does not mean that these objects cannot exist. This issue is considered in section 3. Its analysis is based on the condition of hydrostatic equilibrium, expressed by the Oppenheimer--Volkov equation, 
of an object comprised of a charged anisotropic fluid. Inspection of this equation reveals that, 
as a condition of hydrostatic equilibrium, 
insufficient gravitational binding demands additional binding of nongravitational origin contributed by the stress of the fluid 
forming the object. In the isotropic case, it amounts to the requirement that the fluid must generate sufficient negative pressure, 
at least near the surface of the object, which may or may not be possible depending on the state equation (or equations) for the fluid.  
The Oppenheimer--Volkov equation for a charged anisotropic fluid contains an additional term (compared to the isotropic case) of 
nonrelativistic and nongravitational origin generated by the anisotropy of the stress that may either decrease or increase the total 
binding, though not the gravitational binding. Again, sufficient negative radial pressure, 
tangential pressure, or both are required to compensate for 
insufficient gravitational binding. Any attempt to introduce an object with pure electromagnetic mass by driving the fluid density to zero 
makes it impossible for the object to have sufficient gravitational binding as well as, for any reasonable state equations, non-gravitational 
binding. Such an object simply cannot exist in classical general relativity.  

\section{Gravitational binding of charged spherically symmetric objects}
\label{sec:grbind}
It is well--known \cite{Schutz,MTW} that there is a coordinate system \( t, r, \theta , \phi \) in which the geometry 
of a static spherically symmetric system is given by the line element,

\begin{eqnarray}
ds^{2} = -e^{2\Phi(r)}dt^{2} + e^{2\Lambda(r)}dr^{2} + 
r^{2}(d\theta^{2}+\sin^{2}\theta d\phi^{2}) \nonumber\\
\Rightarrow g_{00} = -e^{2\Phi(r)}; \: g_{rr} = e^{2\Lambda(r)}; \: 
g_{\theta\theta} = r^{2}; \: g_{\phi\phi} = r^{2}\sin^{2}\theta.
\end{eqnarray}

The energy--momentum tensor of a charged perfect fluid is given by

\begin{equation}
T_{\mu\nu} = (\mu + p)\, u_{\mu}u_{\nu} + g_{\mu\nu} p + 
\frac{1}{4\pi}\left({F_{\mu}}^{\kappa}F_{\nu\kappa} - 
\frac{1}{4}F_{\kappa\lambda}F^{\kappa\lambda} g_{\mu\nu}\right) 
\end{equation}

\noindent where \(\mu\) and \( p\) are the proper density and proper pressure of the fluid, respectively.
The notation \(\rho\) is reserved for the proper charge density. The only nonzero component of 
\( u_\mu\) is \(u_0 = -{\rm e}^\Phi\). 

This expression is obtained as the variational derivative of the matter lagrangian with respect to the 
spacetime metric. Although the lagrangian of the charged fluid is composed of the lagrangian of fluid, 
the lagrangian of the electromagnetic field, and the interaction lagrangian (interaction between the charged particles 
of the fluid and the electromagnetic field), the interaction term does not contribute to the energy--momentum. 
The total energy-momentum tensor is the sum of the energy-momentum of the fluid and the energy--momentum 
of the electromagnetic field. This does not mean that the electromagnetic coupling does not contribute 
to the total energy of the object.  Such a contribution, however, is not localizable and appears only 
upon integration over the volume of the object.

The components of the metric tensor and of the energy--momentum tensor allow computation 
of Einstein's equations.  Only four components of these equations are nontrivial and, of those, 
only the 00--component and the rr--component (due to symmetries) contain important physical 
information:

\begin{eqnarray}
G_{00} = 8\pi T_{00} \Rightarrow \frac{1}{r^{2}}e^{2\Phi}\frac{d}{dr}[r(1 - e^{-2\Lambda})] 
= 8\pi \mu e^{2\Phi} + \mathbf{E}^{2}e^{2\Phi}; \\
G_{rr} = 8\pi T_{rr} \Rightarrow - \frac{1}{r^{2}}e^{2\Lambda}(1 - e^{-2\Lambda}) + \frac{2}{r}\frac{d\Phi}{dr} 
= 8\pi p e^{2\Lambda} - \mathbf{E}^{2}e^{2\Lambda}.
\end{eqnarray}

\noindent The other two equations follow from these by virtue of the contracted Bianchi identities. It is 
standard practice to replace them by the equations of motion 

\begin{equation}
{T^{\alpha\beta}}_{;\beta} = 0,
\end{equation}

\noindent where the $ T^{\alpha\beta} $ are components of the 
energy--momentum tensor. With spherical symmetry, the only relevant equation is the one with 
$ \alpha = r $ \cite{Schutz}:

\begin{eqnarray}
{T^{r\beta}}_{;\beta} = {T^{r0}}_{;0} + {T^{rr}}_{;r} + {T^{r\theta}}_{;\theta} + {T^{r\phi}}_{;\phi} = 0 \nonumber\\
\Rightarrow \left( \mu + \frac{E^{2}}{8\pi}\right)\frac{d\Phi}{dr} 
+ \left( p - \frac{E^{2}}{8\pi}\right)\frac{d\Phi}{dr}
+ \frac{d(p - \frac{E^{2}}{8\pi})}{dr}
- \frac{E^{2}}{2\pi r} = 0 \nonumber\\
\Rightarrow \left( \mu + p \right)\frac{d\Phi}{dr}
+ \frac{d(p - \frac{E^{2}}{8\pi})}{dr}
- \frac{E^{2}}{2\pi r} = 0.
\end{eqnarray}

The electric field is constrained by Maxwell's equations which, due to the symmetries of the 
problem, are reduced to the nontrivial equation

\begin{equation}
\nabla\cdot\mathbf{E} = 4\pi\rho
\label{eq:Nontrivial_Maxwell_Equation}
\end{equation}

\noindent where $\rho $ is the proper charge density and $ \mathbf{E} $ 
is the electric field in the proper space of static observers.  Also due to the symmetries, 
$ \mathbf{E} $ is a radial static field, and its magnitude $ E(r) $ can be 
computed by integrating (\ref{eq:Nontrivial_Maxwell_Equation}) with respect to $ r $, yielding

\begin{equation}
E(r) = q(r)/r^{2} 
\end{equation} 

\noindent where $ q(r) $ is the total charge enclosed by a sphere of radius $ r $ 

\begin{equation}
q(r) = \int_{0}^{r}4\pi r^{2}e^{\Lambda(r)}\rho(r)dr.
\end{equation} 

This expression reduces the 00--component of Einstein's equations (referred to hereafter as 
the 00--equation) to 

\begin{equation} 
\frac{d}{dr}\left[ r(1 - e^{-2\Lambda (r)})\right] = 8\pi r^2 \mu (r) + 
\frac{q^2(r)}{r^2} .
\end{equation}  

Subsequent analysis of this equation is described entirely in \cite{CKM}.  We present a slightly 
different account below.  This procedure is of crucial importance for proper understanding of the 
composition of Reissner--Nordstr{\o}m mass.

In this analysis, the boundary of the object \( r = R\) is determined by the conditions \( \mu (r) = \rho (r) = 0\) 
for \( r > R\) which imply that, outside of the object, \(q (r)\) is a constant 

\begin{equation} 
q(r) = q(R) = Q = \int\limits_0^R \rho(r) 4\pi r^2 {\rm e}^{\Lambda (r)} dr 
\end{equation} 

\noindent that can be identified as the total charge of the object.

It is convenient to replace the function \( \Lambda (r)\) by a new function \(m(r)\) 

\begin{equation} 
m(r) = \frac{1}{2} r \left( 1 - e^{-2\Lambda(r)}\right) + \frac{q^2(r)}{2 r}
\end{equation} 

\noindent which reduces the 00--equation to 

\begin{equation} 
\frac{dm}{dr} = 4\pi r^2 \mu (r) + \frac{1}{2 r}\, \frac{d}{dr} [q^2(r)]                       .
\end{equation} 

\noindent or, equivalently

\begin{equation} 
\frac{dm}{dr} = 4\pi r^2 \mu (r) + \frac{q(r)\, \rho (r)\,  4\pi r^2 {\rm e}^{\Lambda (r)}}{r} .
\end{equation} 

Integration of this equation yields \( m(r)\), sometimes called the mass function,  

\begin{equation} 
m(r) = \int\limits_0^r 4\pi r^2 \mu (r) dr + \int\limits_0^r \frac{q(r)\, \rho (r)\, 
4 \pi r^2 {\rm e}^{\Lambda (r)}}{r}\, dr 
\end{equation}

\noindent although it \emph{cannot} be interpreted as the mass--energy inside \( r\) since  
energy is not localizable in general relativity. 

Outside the object ( \( r > R\) ), both integrals become constant, and so does the mass function 

\begin{equation} 
m(r) = m(R) = M = \int\limits_0^R 4\pi r^2 \mu (r) dr + 
\int\limits_0^R \frac{q(r)\, \rho (r)\,  
4 \pi r^2 {\rm e}^{\Lambda (r)}}{r}\, dr .
\end{equation} 

\noindent The constant \( M\) is interpreted as the total mass of the object, since substitution of 

\begin{equation}
e^{2\Lambda(r)} = \frac{1}{1 - \frac{2M}{r} + \frac{Q^{2}}{r^{2}}}, 
\end{equation} 

\noindent in the line element turns it into the Reissner--Nordstr{\o}m metric; this identifies 
\( M\) as the total mass of the object based on analysis of the Keplerian motion of 
neutral test particles around the object \cite{MTW}.  

The second term of this expression represents electric 
coupling of charges in the object. It can be called the \emph{electromagnetic mass} of the object 
(similar to the electromagnetic mass of an electron in the Lorentz theory). 

In the last expression, the second term is an integral over the proper volume whereas 
the first is not. We rewrite the expression for \( M\) in the form 

\begin{equation} 
M = m(R) = \int\limits_0^R {\rm e}^{-\Lambda (r)} \mu (r)\,   4\pi r^2 {\rm e}^{\Lambda (r)} dr + 
\int\limits_0^R \frac{q(r)\, \rho (r)}{r}\,  
4 \pi r^2 {\rm e}^{\Lambda (r)}\, dr . \label{eq:mm}
\end{equation} 

\noindent In (\ref{eq:mm}), integration in both terms is performed over the proper volume, with 
\( 4\pi r^2 {\rm e}^{\Lambda (r)} dr\) the proper volume of the sperical layer between \( r\) 
and \( r + dr\). In the first term, the factor 

\begin{equation} 
{\rm e}^{-\Lambda (r)} = \left[ 1 - \frac{2m(r)}{r} + \frac{q(r)^{2}}{r^{2}}\right]^{\frac{1}{2}} 
\end{equation} 

\noindent contains the contribution of gravitational binding energy to the mass of the object.  
More explicitly, this expression can be written as 

\begin{eqnarray}  
M = m(R) = & \int\limits_0^R  \mu (r)\,   4\pi r^2 {\rm e}^{\Lambda (r)} dr +
\int\limits_0^R \frac{q(r)\, \rho (r)}{r}\,  
4 \pi r^2 {\rm e}^{\Lambda (r)}\, dr \nonumber \\  
& - \int\limits_0^R \left( 1 - {\rm e}^{-\Lambda (r)}\right) \mu (r)\,   4\pi r^2 {\rm e}^{\Lambda (r)} dr,
\label{eq:M_explicit}
\end{eqnarray} 

\noindent where the integral 

\begin{equation} 
\int\limits_0^R \left( 1 - {\rm e}^{-\Lambda (r)}\right) \mu (r)\,   4\pi r^2 {\rm e}^{\Lambda (r)} dr 
\label{eq:M_bind} 
\end{equation} 

\noindent is the gravitational binding energy of the object; its sign is consistent with that in \cite{MTW}.  
It should be noted that the gravitational binding energy is influenced 
by the charge distribution (via the expressions for \( {\rm e}^{-\Lambda (r)}\) and \( m(r)\)) and by 
the pressure to the extent that the pressure influences \( \mu (r)\) through the state equation (or equations). 
The expression for the integral above clearly shows that 
gravity binds only the localized part of the mass (perfect fluid) but not the non--localizable part caused 
by electric coupling. In the case with \( q(r) = 0 \), this integral is positive (we are assuming that 
\( \mu (r)\) is nonnegative) and represents true binding.  
However, when \( q(r) \) is large enough compared to \( m(r) \), the integral might become negative, 
in which case gravity cannot hold the object together.  Removal of all but electromagnetic contributions 
to the mass will produce an object that cannot be held together by gravity. 

In the expression for the total mass \( M\), the second term (electromagnetic mass) includes a contribution from the 
energy of the electric field outside of the object (from \( R\) to \(\infty\)) \cite{Edyn}. That we can write it as 
an integral over the interior of the object (from \( 0\) to \( R\)) is important as it allows us to treat the object 
as isolated \cite{Wald} and to introduce an alternative expression for its mass, the Tolman--Whittaker formula. 
A thorough and modern treatment of the issue was given by Komar \cite{Komar} and M{\o}ller \cite{Mol}. It is applicable 
to the cases we are interested in, namely static (with respect to the timelike Killing vector field \( \xi^\mu\) ), 
asymptotically flat spacetimes and spatially bounded objects (cf. \cite{Wald} for generalizations), which is the case 
of the Schwarzschild and Reissner--Nordstr{\o}m spacetimes. Such a procedure introduces a Tolman--Whittaker version of the 
mass function \( m_G(r)\). It should be noted that neither \( m(r)\) nor \( m_G(r)\) with \( r < R\) produces the value 
of mass for a portion of the object and that \( m(r)\) and \( m_G(r)\), in general, do not coincide. However, according to 
what is sometimes referred to as the general relativistic version of the virial theorem \cite{Wald}, \( m(R) = m_G(R) = M\) 
and does produce the total mass of the object (including the electric field contribution).

It is easy to generalize the results described above to the case of locally anisotropic equations of state for spherically 
symmetric objects. 
An excellent presentation of the issue for the case of neutral objects can be found in \cite{Anis}. These considerations  
have been applied, with different degrees of success, in astrophysics \cite{Anis} and in attempts to produce a more 
successful classical model of an electron \cite{PdeL}.  Local anisotropy together with spherical symmetry implies that the 
energy--momentum tensor \( {T^\mu}_\nu\) (in Schwarzschild coordinates) is diagonal and in the \emph{neutral} case is given by

\begin{equation}
{T^\mu}_\nu = diag\,(-\mu , p_r, p_\theta , p_\phi ), 
\end{equation} 

\noindent with \( p_\theta = p_\phi = p_\perp\), or 

\begin{equation}
{T^\mu}_\nu = diag\,(-\mu , p_r, p_\perp , p_\perp ). 
\end{equation} 
 
Einstein's equations are essentially the same (for both neutral and charged objects) as in the isotropic case, except 
\( p_r\) (radial pressure) replaces \( p\) in the \( rr\)--equation, \( p_\perp\) (tangential pressure) replaces \( p\) in the \(\theta\theta\) 
and \(\phi\phi\) equations and, in general, \( p_r \neq p_\perp\).  Then in the neutral (without charge) case, we obtain three 
structure equations in five unknowns (\(\Phi\), \(\Lambda\), \(\mu\), \( p_r\) and \( p_\perp\)), which implies that, 
in addition, two equations of state \( p_r = p_r(\mu )\) and \( p_\perp = p_\perp (\mu )\) must be specified. 
In general, \( p_r\) and \( p_\perp\) may depend on more variables (such as entropy, etc.) in which case more equations are needed. 
Of course, in the case of a charged object, one more variable \(\rho\) must be included and one more equation should be 
added. A more detailed account of this issue can be found in the literature; we are not going to pursue it further. 

The $00$--equation in the anisotropic case remains the same as in the isotropic case, which implies that the expression for 
the mass function \( m(r)\) and its connection with \(\Lambda (r)\) remain the same for both neutral and 
charged objects, as does the expression for the total mass $ M $ (Reissner--Nordstr{\o}m parameter) of the object. 

We note that attempting to produce an object with purely or mostly electromagnetic mass by making 
\( \mu\) arbitrarily small 
will result in gravitational binding that is both arbitrarily small and insufficient to counter 
the electrostatic repulsion in the object. 

\section{Gravitational binding and total binding}
\label{sec:bianis}

Even if gravitational binding is not strong enough to counter electrostatic repulsion, it does not mean that 
a charged, static spherically symmetric object cannot exist as additional binding can be supplied by the stress of the fluid 
forming the object.
    
The condition of hydrostatic equilibrium in the object -- the pressure gradient needed to keep the 
fluid static in the gravitational and electromagnetic fields \cite{Schutz} -- is described by the Oppenheimer--Volkov (O--V) equation. 
As in the case of a neutral object, the O--V equation is derived by eliminating $ d\Phi/dr $ from 
the rr--component of Einstein's equations and from the equation of motion.  For a charged spherical object, 
it takes the form \cite{Bek}

\begin{equation}
\frac{dp}{dr} = \frac{q(r)}{4\pi r^{4}}\frac{dq(r)}{dr} - \left(\mu +
p\right) \left(\frac{4\pi r^3 p - \frac{q^2(r)}{r} +
m(r)}{r[r-2m(r)+\frac{q^2(r)}{r}]}\right).
\label{eq:ov_charged}
\end{equation} 

\noindent This equation is not separable, unlike the O--V equation for an uncharged spherical object \cite{Schutz} 
obtained from (\ref{eq:ov_charged}) by letting $ q(r) \rightarrow 0 $, 

\begin{equation}
\frac{dp}{dr} = -\frac{(\mu + p)(m(r) + 4\pi r^3 p)}{r(r-2m(r))}.
\label{eq:ov_uncharged}
\end{equation}

\noindent We are assuming that \( \mu > 0\) for \( r < R\). 

The first term on the right hand side of the O--V equation (\ref{eq:ov_charged}) represents electrostatic repulsion, 
does not depend on \( \mu\), and is positive if the sign of \( \rho\) is the same everywhere. The second term supplies 
gravitational binding. If the sum of the two terms is positive, the object cannot be held together by gravitational 
binding alone, and additional binding must be supplied by the stress of the fluid such that (according to the O--V 
equation) \( \frac{dp}{dr} > 0\). The meaning of this requirement is especially transparent near the surface of the object. 
At the surface of the object ($ r = R $), the pressure $ p_{r = R} = 0 $, and the O--V equation reduces to

\begin{eqnarray}
\frac{dp}{dr}|_{r = R} & = & \frac{QQ'}{4\pi R^{4}} - \mu \left(\frac{-\frac{Q^2}{R} +
M}{R[R-2M+\frac{Q^2}{R}]}\right) .
\end{eqnarray} 

Inside the object, the conditions \( \frac{dp}{dr} > 0\) and $ p_{r = R} = 0 $ imply that the pressure of the fluid 
must be negative, at least near the surface of the object. 

The first term of this expression does not depend on \( \mu\) and is positive if the sign of \( \rho\) is the same 
everywhere, which means that when \( \mu\) becomes arbitrarily small, the pressure gradient becomes positive and the pressure of 
the fluid becomes negative no matter the sign of \( \mu\). Of course, a 
configuration with arbitrarily small density and finite negative pressure is thoroughly nonphysical, but we have not 
assigned these quantities by hand:  it is the O--V equation that demands this.  The correct physical interpretation is that such an object 
cannot be formed in classical general relativity.

In the case of an anisotropic fluid,
hydrostatic equilibrium in the object is determined by the anisotropic generalization of the O--V equation 
derived in the same way as the isotropic version. In the absence of charge, it is expressed by

\begin{equation}
\frac{dp_r}{dr} = -(\mu + p_r)\, {\Phi}' + \frac{2}{r}\, (p_\perp - p_r), 
\end{equation} 

\noindent with 

\begin{equation}
{\Phi}' = \frac{m(r) + 4 \pi r^3 p_r}{r\, (r - 2 m)}  
\end{equation} 

\noindent so that, finally, the O--V equation takes the form
 
\begin{equation}
\frac{dp_r}{dr} = -(\mu + p_r)\, \frac{m(r) + 4 \pi r^3 p_r}{r\, (r - 2 m)} + \frac{2}{r}\, (p_\perp - p_r). 
\end{equation} 

\noindent where \( m(r)\) is the mass function as described above. 

It should be noted that there is no equation for the gradient of the tangential pressure \( p_\perp\) as it is determined 
by the equations above plus the equations of state.
 
The second term in the O--V equation is the only one in the structure equations that explicitly contains \( p_\perp\). 
Moreover, in the Newtonian limit the O--V equation reduces to 

\begin{equation}
\frac{dp_r}{dr} = -\frac{m\, \mu}{r^2} + \frac{2}{r}\, (p_\perp - p_r), 
\end{equation} 

\noindent which implies that the anisotropy term is of Newtonian origin in the case of spherical symmetry \cite{Anis}. In addition, we wish 
to point out that this term is produced not by gravity but by the stress of the fluid.

To solve the O--V equation (together with the other structure equations), appropriate boundary conditions must be imposed. 
Just as in the case of isotropy, it is required that the interior of the matter distribution be free of singularities, 
which imposes the condition \( m(r) \rightarrow 0\) as \( r \rightarrow 0\). If \( p_r\) is finite at \( r = 0\), then 
\( {\Phi}' \rightarrow 0\) as \( r \rightarrow 0\). Therefore \(\frac{dp_r}{dr}\) will be finite at \( r = 0\) only if 
\( p_\perp - p_r\) vanishes at least as rapidly as \( r\) when \( r \rightarrow 0\). Ordinarily, it is required that \cite{Anis} 
\begin{equation}
\lim_{r \rightarrow 0} \frac{p_\perp - p_r}{r} = 0. 
\end{equation} 

The radius of the object \( R\) is determined by the condition \( p_r(R) = 0\). It is not required that \( p_\perp(R) = 0\). 
In astrophysical applications, it is assumed that  
\( p_\perp(r) \geq 0\) for all \( r < R\) \cite{Anis}.  
An exterior vacuum metric  (Schwarzschild or, later, Reissner--Nordstr{\o}m) is always matchable to the interior solution 
across \( r = R\) as long as \( p_r(R) = 0\). 

The charged anisotropic O--V equation is obtained as easily (the procedure is the same as for the isotropic case):
\begin{equation}
\frac{dp_r}{dr} = \frac{q(r)}{4\pi r^{4}}\frac{dq(r)}{dr} - \left(\mu +
p_r\right) \left(\frac{4\pi r^3 p_r - \frac{q^2(r)}{r} +
m(r)}{r[r-2m(r)+\frac{q^2(r)}{r}]}\right) + \frac{2}{r}\, (p_\perp - p_r).
\label{eq:ov_chargedan}
\end{equation} 

\noindent All comments concerning the boundary conditions for neutral objects (cf. above) apply to charged objects 
as well. 

At the surface of the object ($ r = R $), the pressure $ p_r = 0 $, so 
\begin{eqnarray}
\frac{dp_r}{dr}|_{r = R} & = & \frac{QQ'}{4\pi R^{4}} - \mu \left(\frac{-\frac{Q^2}{R} +
M}{R[R-2M+\frac{Q^2}{R}]}\right) + \frac{2}{R}\, p_\perp . 
\label{eq:ovsrf}
\end{eqnarray} 

\noindent This equation is the same as the isotropic version except for a term associated with 
tangential pressure. In 
cases when gravity alone is insufficient to counteract electrostatic repulsion, additional binding caused by 
the stress of the fluid is required to satisfy the O--V equation. Near the surface of the object, such stress can be generated 
by anisotropy if \( p_{\perp}(r) < 0\) and, if it is not enough, by allowing \( p_{r}(r) < 0 \). In any case, the O--V equation 
requires one or both of these pressures to be finite (negative). However, for any reasonable state equations (those state equations such that the radial and tangential pressures will become 
arbitrarily small as the density is made arbitrarily small), the O--V equation 
cannot be satisfied for arbitrarily small density. This means that an object with pure or mostly electromagnetic mass 
obtained by making \( \mu\) arbitrarily small cannot exist in classical general relativity. 

\section{Conclusions}

We have investigated gravitational binding of spherically symmetric charged objects comprised of a perfect fluid.
We have shown that the electric coupling term in the mass function is not localizable and does not 
contribute to gravitational binding. Our expression for 
the total mass of a charged object is not meant to help in generating new solutions and does not do so.  
Instead, it provides a physically transparent 
description of the composition of the total mass and allows for drawing important conclusions common 
for all models. In particular, it shows that any attempt to remove all of the mass density of the fluid from 
the object will completely destroy its gravitational binding. 
The generalization of the expressions for the total mass (Reissner--Nordstr{\o}m parameter) and the binding energy 
for the anisotropic case (\( p_r \neq p_\perp\) ) is trivial (formally, they are the same) and does not change the outcome of our analysis 
for the limiting case of objects of pure electromagnetic mass. 

The limitation of analyzing only the expression for the total mass is that the total binding includes a non-gravitational contribution due to the stress of the fluid that does 
not show up explicitly in the expression for the gravitational binding. Its influence can be tracked via inspection of the O--V equation. We have 
done so for an object composed of an anisotropic charged fluid. In cases when gravitational binding alone is insufficient to counter electrostatic 
repulsion, the condition of hydrostatic equilibrium (O--V equation) requires this additional stress to produce a finite negative pressure 
(tangential, radial, or both), at least near the surface of the object. This condition may or may not be possible to 
satisfy, depending on the state equations of the fluid. However, in the particular case of a pure electromagnetic object 
(when \( \mu\) is arbitrarily small), it is impossible to satisfy for any reasonable choice of state equations. This leads to a stronger conclusion than we had before:  
an object with pure electromagnetic mass not only cannot be bound by gravity, but it cannot even exist in classical general relativity.     

We wish to stress that our analysis has been limited to the static spherically symmetric case. It is well--known that, in general, 
there might be gravitationally bound objects of electromagnetic composition (for instance, Wheeler's geon \cite{Whg}).  

\section{Acknowledgements}

We would like to thank the NCSU Department of Mathematics Summer 2009 Research 
Experiences for Undergraduates (REU) Program for support of this work.

\end{document}